\documentclass[aps,prb,twocolumn,superscriptaddress,showpacs]{revtex4}

\renewcommand{\Im}{\text{Im}\,}
\newcommand{\bra}{\langle}
\newcommand{\ket}{\rangle}
\newcommand{\mr}{{\mathbf{r}}}
\newcommand{\mR}{{\mathbf{R}}}
\newcommand{\mN}{{\mathbf{0}}}
\newcommand{\mk}{{\mathbf{k}}}
\newcommand{\mq}{{\mathbf{q}}}
\newcommand{\mb}{{\mathbf{b}}}
\newcommand{\ep}{\varepsilon}
\newcommand{\ph}{\varphi}

\begin{document}

\title{An ab-initio many-body method for electronic structure
  calculations of solids.\\
  I. Description of the method}

\author{I. Schnell}
\affiliation{Theoretical Division,
  Los Alamos National Laboratory, Los Alamos, New Mexico 87545}
\affiliation{Department of Physics, University of Bremen,
  P.O.Box 330 440, D-28334 Bremen, Germany}

\author{G. Czycholl}
\affiliation{Department of Physics, University of Bremen,
  P.O.Box 330 440, D-28334 Bremen, Germany}

\author{R.C. Albers}
\affiliation{Theoretical Division,
  Los Alamos National Laboratory, Los Alamos, New Mexico 87545}

\date{\today}

\begin{abstract}
We propose a new, alternative method for ab-initio calculations of the
electronic structure of solids, which has been specifically adapted to
treat many-body effects in a more rigorous way than many existing
ab-initio methods. We start from a standard
band-structure calculation for an effective one-particle Hamiltonian
approximately describing the material of interest. This yields a
suitable set of one-particle basis functions, from which well localized
Wannier functions can be constructed using a method proposed by Marzari
and Vanderbilt. Within this (minimal) basis of localized Wannier
functions the matrix elements of the non-interacting (one-particle)
Hamiltonian as well as the Coulomb matrix elements can be calculated.
The result is a many-body Hamiltonian in second quantization with
parameters determined from first principles calculations for the
material of interest. The Hamiltonian is in the form of a multi-band
Hamiltonian in second quantization (a kind of extended, multi-band
Hubbard model) such that all the standard many-body methods can be
applied. We explicitly show how this approach can be solved in the 
simplest many-body approximation, the mean-field Hartree-Fock
approximation (HFA), which takes into account exact exchange
and corrects for self-interaction effects.
\end{abstract}


\maketitle

\section{\label{sec:Introduction}Introduction}

Most existing ab-initio (first-principles) methods for the numerical
calculation of the electronic properties of solids are based on density
functional theory (DFT)\cite{HK64}, which in principle is exact and
properly takes into account many-body effects involving the
Coulomb interaction between the electrons; for an overview on the
present status of DFT we refer to the books\cite{DrGross90,Eschrig96}.
But, in general, the functional dependence of the kinetic energy and the
exchange and correlation part of the Coulomb (interaction) energy on the
electron density are not known explicitly, and hence additional
approximations and assumptions are necessary. A well established
additional approximation is the local density approximation
(LDA)\cite{KS65} (or local spin-density approximation, LSDA, for
magnetic systems), which assumes that the electron density of the
interacting system is the same as that of an effective non-interacting
system and that the exchange-correlation potential depends only on the
electronic density locally. Even then, the functional dependence of the
exchange-correlation energy on the density is not known in general, and
it is usually necessary to make an ansatz for the exchange-correlation
functional, which is based on the homogeneous electron gas. The LDA
goes beyond the simplest electron-gas approximation, the
Hartree-Fock approximation (HFA), in that correlation energy (i.e. the
part of the interaction energy beyond HFA) is explicitly taken into
account.  More recent generalized gradient approximation (GGA)
calculations also make some improvements in the correlation energy that
slightly improve on the homogeneous electron gas approximation. On the
other hand, the exact HFA exchange potential is non-local, i.e., it
depends on the electron wave functions and density at all other
positions, an effect which the local LDA exchange potential misses.
Also, LDA misses some cancellation between the exchange and correlation
parts of the total energy (self-interaction corrections). However, in
practice LDA-treatments are simpler than HFA-calculations (especially
for metals), because local exchange is easier to treat than non-local
exchange, and are usually in better agreement with experiment.
Therefore, DFT-(LDA-like) treatments have been far more common than HFA
during the past few decades, even in quantum chemistry (with a long
tradition of methods based on HFA).

Ab-initio DFT calculations based on the L(S)DA have been very successful
for many materials and ground-state properties such as crystal
structure, ground state and ionization energy, lattice constant, bulk
modulus, crystal anharmonicity\cite{Rose84}, magnetic moments, and some
photo emission spectra. However, there are also important limitations.
For example, LDA predicts a band gap for semiconductors that is almost a
factor of two too small. In addition, for many strongly correlated
(narrow energy band) systems such as high-temperature superconductors,
heavy fermion materials, transition-metal oxides, and $3d$ itinerant
magnets, the LDA is usually not sufficient for an accurate description
(predicting metallic rather than semiconducting behavior, failing to
predict quasi-atomic-like sattelites, etc).

Therefore, it is justified and important to look for new, better
ab-initio methods and improvements that go beyond L(S)DA. In fact, there
have already been several attempts to improve LDA, such as gradient
corrections, non-local density schemes, self-interaction corrections,
and the GW approximation. Gradient corrections\cite{DrGrossKap7}
approximately account for the fact that the electron density is not
constant but $\mr$-dependent in an inhomogeneous electron gas; estimates
on the influence of perturbations of the homogeneity on the ground state
energy of the homogeneous electron gas are then used to construct a new
functional for the exchange-correlation potential containing $\nabla
n(\mr)$ terms. The non-local density schemes go beyond LDA by
considering that the exact exchange-correlation potential
$V_{\text{xc}}(\mr)$ cannot depend only on the density $n(\mr)$ at the
same position $\mr$ but should depend also on the electron density at
all other positions $n(\mr')$. Usually the new ansatz for the functional
of the exchange-correlation energy contains the pair correlation
function or the interaction of the electrons with the
exchange-correlation hole.\cite{Gunnarsson80,DrGrossKap7}. The recently
developed exact exchange (EXX) formalism\cite{Vogl97} cancels the
spurious (unphysical) electronic self-interaction present in LDA and
gradient corrected exchange functionals. A standard method for ab-initio
calculations of excited states is the GW-approximation
(GWA)\cite{Hedin,Aulbur}. Denoting the one-particle Green function by
$G$ and the screened interaction by $W$, the GWA is an approximation for
the electronic selfenergy $\Sigma \approx GW$, which is correct in
linear order in $W$ and can diagramatically be represented by the
lowest-order exchange (Fock) diagram. The one-particle Green function
$G$ is usually obtained for the effective one-particle LDA
Hamiltonian.

Recently there have been several attempts to combine ab-initio LDA
calculations with many-body
approximations.\cite{AZA91,SASS91,SAS9294,APKAK97,LK98,DJK99,KL99,LL00,
WPN00,NHBPAV00} All of these recent developments add local, screened
Coulomb (Hubbard) energies U between localized orbitals to the
one-particle part of the Hamiltonian obtained from an ab-initio LDA
band-structure calculation, but differ in how they handle the
correlation part. In the earliest attempts, the LDA+U method\cite{AZA91}
used essentially a static mean-field-like (or Hubbard-I-like)
approximation for the correlation.  The simplest approximation beyond
Hartree-Fock, second-order perturbation theory (SOPT) in U, was
used\cite{SASS91,SAS9294,DJK99,WPN00} to study the electronic properties
of 3d-systems (like Fe and Ni) and heavy fermion systems (like UPt$_3$).
The LDA++ approach\cite{LK98,KL99,LL00} has a similar strategy, but uses
other many-body approximations to treat the correlation problem, namely,
either the fluctuation-exchange approximation (FLEX) or the dynamical
mean-field theory\cite{GKKR96} (DMFT); for a review on this recently
very successful LDA+DMFT\cite{NHBPAV00} approach see Ref.
\onlinecite{HeldVollhardtetal}. These approaches, including the LDA+U,
have in common that they have to introduce a Hubbard U as an additional
parameter and hence are not really first-principles (ab-initio)
treatments.  Although they use an LDA ab-initio method to obtain a
realistic band structure, i.e., single-particle properties, Coulomb
matrix elements for any particular material are not known, and the
Hubbard U remains an adjustable parameter. In addition, since some
correlations are included in LDA as well as by the Hubbard U, it
is unclear how to separate the two effects and double-counting of
correlation may be included in these approximations.

In this paper we suggest a different, alternative approach and propose a
new type of ab-initio treatment. Because we want to avoid DFT-LDA due to
its problems with double counting and self-interactions, we instead
suggest starting from a realistic Hamiltonian in second quantization and
directly applying many-body methods to this Hamiltonian. The
underlying many-body theory in DFT-LDA involves the homogeneous electron
gas, from which a functional for the exchange-correlation potential is
derived. But we want to apply the many-body theory to a Hamiltonian tailored
to the material of interest, i.e., study a
problem of interacting electrons on a lattice and therefore an
inhomogeneous electron gas. To get the realistic, microscopic
Hamiltonian in second quantization the one-particle and Coulomb matrix
elements have to be calculated. For that purpose one needs a suitable
one-particle basis. 
For a solid it is
reasonable that either Bloch or Wannier functions should form a suitable
one-particle basis set. We suggest starting from a set of maximally
localized first-principles Wannier functions, because the good
localization of these wave function should make the one-particle
(tight-binding) and two-particle (Coulomb) matrix elements important
only for a finite number of near-neighbor shells (on-site, nearest, next
nearest neighbors, etc.) such that only a few matrix elements have to be
actually calculated. 
It is possible to determine these Wannier states for an
effective (auxiliary) one-particle Hamiltonian. For example, 
one could use the true
non-interacting Hamiltonian, the Hartree Hamiltonian, or even the LDA
Hamiltonian. 
In principle, the final results should not depend on the
choice of this auxiliary one-particle Hamiltonian, because any set of
Wannier functions should span the whole one-particle Hilbert space. But,
in practice, one typically works only
within a restricted (finite dimensional) subspace. 
Therefore, the auxiliary one-particle
Hamiltonian and the resulting one-particle basis need to be chosen in
such a way that the physically most important  subspace is
spanned. In practice, we have found the Hartree Hamiltonian to be a
suitable choice for the auxiliary, effective one-particle Hamiltonian.

Our suggested procedure is schematically outlined as follows:
First, a traditional band-structure calculation is used for the
auxiliary one-particle Hamiltonian to calculate the
eigenstates in the form of Bloch functions. Wannier functions are
closely related to these Bloch functions via a unitary transformation,
and thus span the same one-particle space as the Bloch functions.
However, since the phases of the Bloch functions are arbitrary, Wannier
functions are not unique. Their non-uniqueness (gauge freedom) is
used to construct ``maximally localized Wannier functions'' using a
method proposed by Marzari and Vanderbilt\cite{MV97}. A proper
localization of the Wannier functions is important, because only then do
the standard assumptions that are frequently used in model treatments hold,
e.g., that both one-particle (tight-binding) and two-particle (Coulomb)
matrix elements are important only on-site and for a few neighbor
shells. Next, in order to describe the physical Hamiltonian in second
quantization, one has to evaluate a restricted number of hopping and
Coulomb matrix elements within the basis of the well localized Wannier
functions. Making use of the special representation of the wave
functions inherent to the specific band-structure method used (for
instance in terms of spherical harmonics, plane waves or Gaussian
orbitals) can simplify the actual computation of these matrix elements;
for the Coulomb matrix elements we propose, in particular, an algorithm
based on fast Fourier transformation, which works independent of the
special band-structure method. The result of this procedure is a
multi-band, second-quantized Hamiltonian in a Wannier representation
(a kind of extended, multi-band Hubbard model) describing interacting
electrons on a lattice.  Unlike model studies of correlated lattice
electrons, the matrix elements of this Hamiltonian are not free
adjustable parameters, but are obtained from first principles for the
given material. This Hamiltonian should be evaluated by an
appropriate many-body methods. The simplest approximation is the 
mean-field  Hartree-Fock
approximation (HFA), which exactly treats exchange and
excludes self-interaction effects. Unlike first-quantization treatments
of HFA, which require the solution of a Schr{\"o}dinger
equation with a non-local potential, in second-quantization
only the selfconsistent evaluation of expectation values is required 
(once an appropriate one-particle basis has been chosen).
But, of course, HFA
is certainly not sufficient for realistic materials properties; higher
order correlations have to be taken into account, for example,
perturbationally by resummations of Feynman diagrams (RPA-bubble
diagrams leading to screening, ladder diagrams, etc.) or employing
recently developed non-perturbational methods like dynamical mean-field
theory (DMFT) (if the basic assumptions, for instance concerning the
locality of the most important interaction terms, hold for the specific
many-body Hamiltonian).

This paper is organized as follows: In Section \ref{sec:Hamiltonian} we
shortly repeat some basic notations, give the Hamiltonian in first and
second quantization, describe DFT and LDA, Wannier- and Bloch-states as
possible basis states spanning the one-particle Hilbert space and the
necessary truncation.
Section \ref{sec:bandstructure} describes how a suitable
restricted basis set is obtained by solving the Schr{\"o}dinger equation
for an auxiliary one-particle Hamiltonian, which should be chosen
appropriate for the material of interest and incorporating information
about the lattice structure, lattice constant, total electron number,
etc. In Section \ref{sec:Wannierfunctions} we discuss how another
equivalent basis set of maximally localized Wannier functions can be
obtained from this one-particle basis. Within this Wannier basis
ab-initio one-particle (tight-binding) and two-particle (Coulomb) matrix
elements can be calculated, as outlined in section \ref{sec:matrixel}.
The resulting many-body Hamiltonian in second quantization with
ab-initio parameters is then ready to be studied by suitable many-body
approximations, which is the subject of Section \ref{sec:manybody}. We
then discuss and compare the advantages and disadvantages of our
proposed ab-initio method relative to other ab-initio methods in Section
\ref{sec:comparison}. The following paper (II) contains an initial
simple application (an unscreened HFA) of this approach to the 3d
transition metals Fe, Co, Ni and Cu.

\section{\label{sec:Hamiltonian}Hamilton Operator in first and second
quantization}

A system of $N_e$ interacting (non-relativistic) electrons can be
described by the Hamiltonian
\begin{equation} \label{eq:firstquant}
  H = T + V + W =
  \sum_{i=1}^{N_e} \frac{\mathbf{p}_i^2}{2m} +
  \sum_{i=1}^{N_e} V(\mr_i) +
  \sum_{i > j} \frac{e^2}{|\mr_i-\mr_j|} ~~.
\end{equation}
The first part $T$ is the kinetic energy of the electrons. The $V(\mr)$
describes the external one-particle potential, which for molecules and
solids is solely caused by the positive nuclei or ions and--within the
adiabatic or Born-Oppenheimer approximation--can be regarded as a static
potential. For solids, i.e., for electrons in a crystal, $V(\mr)$ is a
periodic potential. The third part $W$ is the Coulomb interaction
between the electrons. 

For the interacting system it is impossible to exactly solve the full
Schr{\"o}dinger equation and to calculate the many-body eigenenergies
(or even the ground state energy) and the totally antisymmetric
many-particle wave functions, which can be represented as a linear
combination of Slater determinants (a configuration-interaction or CI
approach to correlation). This is practical only for a very small number
of electrons. The basic idea of DFT is to avoid determining the
antisymmetric (ground state) wave function and instead to focus on the
(ground state) density $n(\mr)$. This requires knowing the functional
dependence of the kinetic energy $T$, the one-particle potential $V$,
and the interaction energy $W$ on the density $n$. From this the ground
state density can be determined from a variational principle. But in
general this DFT-concept can only be approximately applied, since the
functional dependence of the kinetic energy and the exchange-correlation
energy on the density is unknown. Therefore additional assumptions and
approximations are necessary, such as the LDA  where the
density $n(\mr)$ is the same as a non-interacting electron system in an
effective one-particle potential and the exchange-correlation energy
depends only locally on the electron density. From a fundamental point
of view these additional assumptions are hard to justify and
uncontrolled, and an estimate of the error or systematic corrections are
very difficult to obtain.

In this paper we choose yet another approach that employs the (very
elegant) formalism of ``second quantization'', which automatically
accounts for the antisymmetry through the fermion anticommutation
relations. In second quantization the full many-body Hamiltonian can be
written as:
\begin{equation} \label{eq:secondquant}
  H = \sum_{i,j,\sigma} t_{ij} c_{i\sigma}^{\dagger} c_{j\sigma} + 
  \frac{1}{2} \sum_{i,j,k,l,\sigma,\sigma'}
  W_{ij,kl} c_{i\sigma}^{\dagger}c_{j\sigma'}^{\dagger}c_{k\sigma'}c_{l\sigma} 
\end{equation}
Here $i,j,k,l$ denote a complete set of one-particle orbital quantum
numbers and $\sigma,\sigma'$ are the spin quantum numbers.  The states
$|i\ket$ and the corresponding wave functions $\ph_i(\mr)=\bra\mr|i\ket$
form a basis of the one-particle Hilbert space. The matrix elements in
Eq. \ref{eq:secondquant} are defined by:
\begin{eqnarray} \nonumber
  t_{ij} &=& \bra i| \frac{{\bf p}^2}{2m} + V(\mr) | j \ket
\\ \nonumber
         &=& \int d^3r~ \ph_i^*(\mr)
             \Big( -\frac{\hbar^2}{2m}\nabla^2 + V(\mr) \Big) \ph_j(\mr)
\\ \label{eq:matrixel}
  W_{ij,kl} &=& \bra i|\bra j| \frac{e^2}{|\mr-\mr'|} |k\ket|l\ket
\\ \nonumber
            &=& \int d^3r \int d^3r'
                \ph_{i}^*(\mr) \ph_{j}^*(\mr') \frac{e^2}{|\mr-\mr'|}
                \ph_{k}(\mr')\ph_{l}(\mr)
\end{eqnarray}
Of course these matrix elements depend on the one-particle basis
$\{|i\ket\}$ that is chosen.  
But because of the completeness relation the physical results should, in
principle, not depend on the choice of the one-particle basis. 
Because of the lattice
periodicity an obvious choice for a one-particle basis is a Bloch basis
$\{|n\mk\ket\}$; then the orbital one-particle quantum numbers $n,\mk$
are the band index $n$ and the wave number $\mk$ (within the first
Brillouin zone). The Bloch wave functions can be written as a product of
a plane wave with a lattice periodic Bloch factor
\begin{eqnarray} \label{eq:Blochcondition}
  \psi_{n\mk}(\mr) &=& \bra\mr|n\mk\ket = e^{i\mk\mr} u_{n\mk}(\mr)
\\ \nonumber
  u_{n\mk}(\mr) &=& u_{n\mk}(\mr+\mR)
\end{eqnarray}
where $\mR$ denotes a lattice vector. 
In practice one can work only on a truncated, finite-diemnsional
one-particle Hilbert space. Here the truncation consists of including 
only a finite number of bands and a set of
$\mk$-values from a discrete mesh in $\mk$-space. 
A Bloch basis can be obtained by applying a traditional band-structure
method to solve the Schr{\"o}dinger equation for an effective one-particle
Hamiltonian with a periodic potential.
But, because the Bloch
states are delocalized, a very large number of Coulomb matrix elements
(depending on four quantum numbers) between all possible $\mk$-states
would have to be calculated. Therefore, it seems that a more
appropriate basis would be to use well localized wave functions.
Although the one-particle part is not diagonal within such a localized
basis, it is expected that a short-range tight-binding assumption will
hold, i.e., that the on-site and the intersite matrix elements for only
a few neighbor shells are sufficient. It is well known from elementary
solid state theory that Wannier states provide an alternative
orthonormal basis set, which spans the same space as the Bloch states.
The Wannier states are related to the Bloch states by the unitary
transformations:
\begin{eqnarray} \label{eq:wannier1}
  w_{\mR n}(\mr) &=& \bra \mr|\mR n\ket
  = \frac{1}{N} \sum_{\mk} e^{-i\mk\mR}\psi_{n\mk}(\mr)
\\ \nonumber
  |\psi_{n\mk}\ket &=& \sum_{\mR} e^{i\mk\mR} |\mR n\ket
\end{eqnarray}
Now our strategy is the following:
\begin{itemize}
\item Perform a traditional band-structure calculation for an effective
  one-particle Hamiltonian $H_{\text{eff}}$ with lattice periodicity
  to obtain a Bloch basis of the Hilbert space.
  Only a finite number of band indices will be considered and
  the calculations will be done for a discretized,
  finite mesh in $\mk$-space, i.e., we will work only on a
  reduced, truncated Hilbert space.
\item Determine well-localized Wannier functions spanning the same
  (truncated) Hilbert space as the Bloch basis from the
  canonical transformation (\ref{eq:wannier1}) described above.
\end{itemize}
Of course, the important energy bands (and corresponding band indices)
are those that determine the electronic properties of the system, i.e.,
the bands near to the Fermi level. Because the Hilbert space is
truncated, we do no longer work with a complete basis set. Hence, it is
important to start from Bloch functions obtained from a band-structure
calculation for a well chosen effective one-particle Hamiltonian.
Possible and suitable choices for the effective (auxiliary) one-particle
Hamiltonian will be discussed in the next Section
\ref{sec:bandstructure}. Furthermore, the Wannier functions obtained
according to (\ref{eq:wannier1}) are not unique, because the phases of
the Bloch functions can be chosen arbitrarily. In fact, for a given set
of randomly chosen phase factors the Wannier functions may not be
localized at all. On the other hand, one may use this gauge freedom to
construct optimally localized Wannier functions\cite{MV97}, as will be
described in Section \ref{sec:Wannierfunctions}.

\section{\label{sec:bandstructure}Band structure calculation for
the auxiliary one-particle problem}

Our Bloch state basis should be suitably chosen for the
material of interest, and should be obtained from a standard
band-structure calculation of an effective one-particle
Hamiltonian with a lattice periodic potential:
\begin{equation}
  V_{\text{eff}}(\mr) = V_{\text{eff}}(\mr + \mR)
\end{equation}
for any lattice vector $\mR$. The corresponding
one-particle Schr{\"o}dinger equation is
\begin{equation}\label{eq:Schroedinger}
  \Big(\frac{{\bf p}^2}{2m} + V_{\text{eff}}(\mr)\Big)\psi_{n\mk}(\mr)
  = \ep_{n}(\mk) \psi_{n\mk}(\mr) ~~.
\end{equation}
Different choices for the effective periodic potential are possible. The
simplest choice would be the bare one-particle potential $V(\mr)$.
However, since the Coulomb interaction is strong, in general, the
resulting eigenenergies of this non-interacting Hamiltonian will be much
lower than the relevant energies of the interacting system. For example,
in the 3d-systems like Ni or Cu the 3d- and 4s-states form bands close
to the Fermi energy. However, without any Coulomb repulsion the
3d-states become very strongly bound atomic-like (core) states, which
would be pushed well below the Fermi energy, and therefore the
corresponding Bloch eigenfunctions are not a good starting point to
describe the electronic bands close to the Fermi level. Because the
Hilbert space is truncated, it is extremely important to start from a
band Hamiltonian $T+V_{\text{eff}}$ that gives eigenfunctions as close as
possible to those which are expected to form the relevant many-body states of
the interacting system.
The bare one-particle potential is consequently a bad choice.

Another possible choice for the effective periodic one-particle
Hamiltonian is the Hartree Hamiltonian. Then the Bloch basis is obtained
by solving the one-particle Schr{\"o}dinger equation
\begin{equation} \label{eq:Hartree1quant}
  \Big(\frac{{\bf p}^2}{2m} + V(\mr) +
  V_{\text{H}}(\mr)\Big)\psi_{n\mk}(\mr)
  = \ep_{n}(\mk) \psi_{n\mk}(\mr)
\end{equation}
where the Hartree potential is given by
\begin{equation} \label{eq:Hartreepotential}
  V_{\text{H}}(\mr) = \int d^3r' \frac{e^2 \rho(\mr')}{|\mr - \mr'|}
\end{equation}
and the particle density
\begin{equation} \nonumber
  \rho(\mr) = \sum_{l\mk\sigma} f(\ep_l(\mk)) |\psi_{l\mk}(\mr)|^2
\end{equation}
\begin{equation} \nonumber
  \text{where~~~} f(E) = \frac{1}{e^{\beta(E-\mu)}+1}
\end{equation}
is the Fermi function (step function for the ground state $T=0$ or
$\beta=\infty$). Since the occupied states $|\psi_{l\mk}\ket$ determine
the particle density and Hartree potential, Eq. \ref{eq:Hartree1quant}
has to be solved selfconsistently, which is usually done by iteration
procedure. The advantage of including the Hartree potential into the
effective potential, i.e. $V_{\text{eff}}(\mr) = V(\mr) +
V_{\text{H}}(\mr)$, is that the approximate effects of the Coulomb
interaction are included (in the mean-field approximation). Therefore,
the eigenenergies (energy bands) will be about the right magnitude and
the resulting basis functions can be expected to be more suitable in the
energy regime around the Fermi level. 

Since the only purpose in solving the effective one-particle
Schr{\"o}dinger equation (\ref{eq:Schroedinger}) is the construction of
a suitable basis set of Bloch functions, we will not make use of the
eigenenergies $\ep_{n}(\mk)$ obtained in (\ref{eq:Schroedinger}) or give
these solutions any physical interpretation. For this reason one could
also use a variety of different artificial effective potentials. For
example, we have studied a ``weighted Hartree potential'' 
\begin{equation}
  V_{\text{eff}}(\mr) = V(\mr) + x V_{\text{H}}(\mr)
\end{equation}
with an additional parameter $x$, which can be varied so that the choice
of the basis functions minimizes the resulting Hartree- or Hartree-Fock
ground state energy. Another possible choice for the effective
(auxiliary) one-particle potential would be the LDA-potential.
Because, in this context, LDA would only be used to construct basis
functions, there would be no problem with double counting
interaction terms.

After choosing the appropriate effective one-particle potential one has
to employ a numerical method to solve the Schr{\"o}dinger equation
(\ref{eq:Schroedinger}) for the periodic potential. For this established
and well known band-structure methods can be used, such as the
``linearized muffin-tin orbital'' (LMTO)
method\cite{OKAndersen,Skriver}, the (full potential) ``linearized
augmented plane wave'' (LAPW) method (using the WIEN2k computer
code\cite{WIEN2K}), or the ``full-potential local-orbital''
(FPLO)\cite{Eschrig99} minimum-basis method.

\section{\label{sec:Wannierfunctions}Determination
of maximally localized Wannier functions}

The band-structure calculation diagonalizes the auxiliary one-particle
Hamiltonian to determine the Bloch states $\psi_{n\mk}(\mr) =
e^{i\mk\mr} u_{n\mk}(\mr)$ labeled by band indices $\{n\}$. After
truncating this one-particle Hilbert space by considering only a finite
number $J$ of band indices, unitary transformations are possible for
each fixed $\mk$ that lead to a new basis set
\begin{equation} \label{eq:gauge}
  \tilde{\psi}_{n\mk}(\mr) = \sum_m U_{mn}^{\mk} \psi_{m\mk}(\mr) ~~,
\end{equation}
which are still Bloch functions but with other band indices (within
which the auxiliary one-particle Hamiltonian is, in general, no longer
diagonal). One can use this gauge freedom to construct a Bloch basis
(still of the same restricted one-particle Hilbert space) for which the
corresponding Wannier functions determined according to
(\ref{eq:wannier1}) are maximally localized.

A suitable measure for the localization of the Wannier function is the
spread functional 
\begin{equation} \label{eq:om}
  \Omega = \sum_n \left[ \langle r^2 \rangle_n -
                         \langle \mr \rangle^2_n \right] ~~,
\end{equation}
where the notation $\langle A\rangle_n=\langle\mN n\vert A\vert\mN n
\rangle$ for any operator $A$ has been used. Among all the possible
gauges, we will use that gauge which minimizes this spread functional
$\Omega$. 

A method for minimizing Eq.~(\ref{eq:om}) has been developed by Marzari
and Vanderbilt\cite{MV97}. This method has already found widespread
applications recently\cite{MV2002,Scalettar2002}. It starts from a
decomposition into invariant, diagonal, and off-diagonal contributions:
\begin{eqnarray} \nonumber
  \Omega &=& \Omega_{\text{I}} + \widetilde{\Omega}
\\ \label{eq:om2}
  &=& \sum_n \left[ \bra r^2 \ket_n
     - \sum_{\mR m} \Big| \bra \mR m|\mr|\mN n \ket \Big|^2 \right]
\\ \nonumber
  &+& \sum_n \sum_{\mR m \ne \mN n}
     \Big| \bra \mR m|\mr|\mN n \ket \Big|^2
 \end{eqnarray}
The first term is ``gauge-invariant'', i.e., independent of the choice
of unitary transformations among the bands.  The second term can be
decomposed into band-off-diagonal and band-diagonal components:
\begin{equation} \label{eq:om3}
  \widetilde{\Omega} =
  \underbrace{ \sum_{m\ne n} \sum_{\mR}
    \Big| \bra \mR m|\mr|\mN n\ket \Big|^2
  }_{\mbox{$\Omega_{\text{OD}}$}} +
  \underbrace{ \sum_n \sum_{\mR\ne\mN}
    \Big| \bra \mR n|\mr|\mN n\ket \Big|^2
  }_{\mbox{$\Omega_{\text{D}}$}}
\end{equation}
In practice one needs the unitary matrices $U_{mn}^{\mk}$ on a
discretized $\mk$-mesh, which can be chosen to be simple cubic. If $\mb$
denotes the vectors connecting each $\mk$-point to its nearest
neighbors, one can define
\begin{equation} \label{eq:M}
  M_{mn}^{\mk,\mb} = \bra u_{m\mk} | u_{n,\mk+\mb} \ket ~~,
\end{equation}
where $u_{n\mk}$ are the Bloch factors appearing in
(\ref{eq:Blochcondition}). Using (\ref{eq:M}) one can express the
expectation values of $\mr$ and $r^2$ as\cite{MV97}:
\begin{eqnarray} \label{eq:mv31}
  \bra \mr \ket_n &=& - \frac{1}{N} \sum_{\mk,\mb} w_b \mb \,
  \Im \ln M_{nn}^{\mk,\mb} ~~ w_b=\frac{1}{2b^2}
\\ \label{eq:mv32}
  \bra r^2 \ket_n &=& \frac{1}{N} \sum_{\mk,\mb} w_b
  \left\{ [ 1 - |M_{nn}^{\mk,\mb}|^2 ] + [ \Im \ln M_{nn}^{\mk,\mb} ]^2
\right\} \nonumber
\end{eqnarray}
These expressions are not unique.  This non-uniqueness arises from the
fact that it is {\it only} required that these expectation values become
exact in the limit of a dense mesh. The present choice guarantees that
under the gauge transformation
$|u_{n\mk}\ket \to e^{-i \mk \mR} | u_{n\mk} \ket$
(corresponding to a translation by a lattice vector)
the following physical properties are fulfilled:
\begin{eqnarray} \label{eq:mv24}
  \bra \mr \ket_n  & \to & \bra \mr \ket_n + \mR
\\ \nonumber
  \bra r^2 \ket_n  & \to & \bra r^2 \ket_n + 2 \bra \mr \ket_n \mR + R^2 ~~,
\end{eqnarray}
Then one finds for
$\Omega=\Omega_{\text{I}}+\Omega_{\text{OD}}+\Omega_{\text{D}}$:
\begin{eqnarray} \label{eq:mv34}
  \Omega_{\text{I}}
     &=& \frac{1}{N} \sum_{\mk,\mb} w_b \left(
         J - \sum_{mn} |M_{mn}^{\mk,\mb}|^2 \right)
\\ \nonumber
  \Omega_{\text{OD}}
     &=& \frac{1}{N} \sum_{\mk,\mb} w_b \sum_{m \ne n} |M_{mn}^{\mk,\mb}|^2
\\ \nonumber
  \Omega_{\text{D}}
     &=& \frac{1}{N} \sum_{\mk,\mb} w_b \sum_n \left(
    - \Im \ln M_{nn}^{\mk,\mb} - \mb \bra \mr \ket_n \right)^2
\end{eqnarray}
The Marzari-Vanderbilt algorithm\cite{MV97} now consists in an iterated
application of a (in principle infinitesimal, in practice numerically
discrete) canonical transformation chosen so that the gradient of
$\Omega$ is negative, i.e., the spread functional and thus the
delocalization of the Wannier function decreases. In each iteration
cycle $\Omega_{\text{I}}$ should not change whereas
$\Omega_{\text{OD}}+\Omega_{\text{D}}$ should decrease.

In practice the iteration cycle can schematically be written down
as\cite{MV97}:
\begin{equation} \label{eq:mv60}
  U^{(N)\mk} = U^{(N-1)\mk} ~
  \exp \left[ \Delta W^\mk( M^{(N-1)\mk,\mb} ) \right]
\end{equation}
\begin{equation} \label{eq:mv61}
  M^{(N)\mk,\mb} = U^{(N)\mk ~ \dag} ~ M^{(N-1)\mk,\mb} ~ U^{(N)\mk+\mb}
\end{equation}
with the initialization:
\begin{equation}
  U^{(0)\mk}_{mn} = \delta_{mn}
\end{equation}
\begin{equation} \label{eq:mv58}
  M^{(0)\mk,\mb}_{mn} = \bra u_{m\mk} | u_{n,\mk+\mb} \ket
\end{equation}
Here $\Delta W^\mk(M^{\mk,\mb})$ is an anti-hermitian matrix (so that 
$\exp \left[ \Delta W^\mk \right]$ is unitary) which according to
Ref. \onlinecite{MV97} is explicitly given by the following equations:
\begin{equation} \label{eq:DeltaW}
  \Delta W^\mk = \frac{\alpha}{w} \sum_{\mb} w_b \left({\cal A}(R^{\mk\mb}) -
  {\cal S}(T^{\mk\mb})\right)
\end{equation}
where ${\cal A}(B) = (B- B^\dagger)/2 \, , {\cal S}(B) = (B+B^\dagger)/2i$
are the symmetrizing and antisymmetrizing operations for the operators or
matrices $B$, $w= \sum_{\mb} w_b$, $\alpha \in [0,1]$ is a numerical parameter
determining the discrete, finite step size, and the $J \times J$-matrices
$R^{\mk\mb}$ and $T^{\mk\mb}$ are explicitly given by:
\begin{eqnarray}
  R^{\mk\mb}_{mn} &=& M^{\mk\mb}_{mn}M^{\mk\mb*}_{nn}
\\ \nonumber
  T^{\mk\mb}_{mn} &=&
  \frac{M^{\mk\mb}_{mn}}{M^{\mk\mb}_{nn}}\left(\mbox{Im}\ln M^{\mk\mb}_{nn} +
  \mb \bra \mr \ket_n\right)
\end{eqnarray}  
The whole algorithm (\ref{eq:mv60}) can be considered as an iteration
scheme to construct the unitary transformation $U^\mk$. Once this
iteration has converged one can perform the transformation to the new
Bloch functions according to (\ref{eq:gauge}) and then calculate the
optimally localized Wannier functions according to the definition
(\ref{eq:wannier1}).

In practice it is useful to prepare the Bloch orbitals to make the
starting Wannier functions somewhat localized. This has two advantages:
(i) the minimization procedure converges faster, and (ii) this helps to
avoid getting trapped in local minima. For that purpose we make a rather
trivial ``gauge transformation'' to each Bloch function obtained by the
band structure calculation, namely a multiplication with a phase factor
according to
\begin{equation} \label{eq:il}
  \psi_{n\mk}(\mr) ~~\to~~
  \exp \big[ -i\,\Im\ln \psi_{n\mk}(\mr_n) \big]
  \psi_{n\mk}(\mr) ~~.
\end{equation}  
This gauge transformation has the property that
$\Im\ln\Psi_{n\mk}(\mr_n)$ transforms to zero. So at the point $\mr_n$,
all the Bloch functions will have the same phase and $\bra \mr_n| \mN n
\ket$ will take a large value. To make the method work well, one should
choose $\mr_n$ where the Wannier functions are expected to be reasonably
large and one can choose $\mr_n$ individually for each band $n$.

\section{\label{sec:matrixel}One-particle and Coulomb matrix elements}

After the maximally localized Wannier functions $w_{\mR n}(\mr)$ have
been determined, the next task is to calculate the one-particle and
Coulomb matrix elements of the Hamiltonian. The one-particle matrix
elements in Wannier representation are explicitly given by:
\begin{equation}\label{eq:onepartWan}
  t_{12} = \int d^3r w_{1}^*(\mr)
  \left(-\frac{\hbar^2}{2m}\nabla^2 + V(\mr)\right)w_{2}(\mr)
\end{equation}
where $V(\mr)$ is the lattice periodic one-particle potential for the
non-interacting electrons.
Here, and in the following, we use the abbreviated notation
1 to mean $\mR_1n_1$ and 2 to mean for $\mR_2n_2$, etc.
Note that
the one-particle part of the Hamiltonian cannot be expected to be band
diagonal because of the unitary transformation to another set of the
band indices within which the Wannier functions are maximally localized.
But by construction and definition the Wannier functions obey the
following translational invariance property:
\begin{equation} \label{eq:wtrans}
  w_{\mR n}(\mr) = w_{\mN n}(\mr - \mR)
\end{equation}
where $\mN$ is the arbitrarily chosen zero lattice vector.
Thus, the one-particle matrix elements depend only on the relative
distance $\mR=\mR_1-\mR_2$:
\begin{equation} \label{eq:hopping}
  t_{\mR n,m} = \bra \mR n|\left(-\frac{\hbar^2}{2m}\nabla^2
  + V(\mr)\right) |\mN m \ket
\end{equation}
Because of the strong localization of the Wannier functions one can
safely expect that these one-particle (hopping) matrix elements decrease
with increasing distance $|\mR|$ and have to be evaluated explicitly
only for a limited number of $\mR$, in particular on-site, i.e. for
$\mN$, and for a few neighbor shells $\mR$. In practice it turned out
that for properly localized Wannier functions (optimized according to
the Marzari-Vanderbilt algorithm) the explicit evaluation of
one-particle matrix elements up to the 5th neighbor shell is sufficient,
whereas for non-optimally-localized Wannier functions (made only
``somewhat localized''  according to the prescription (\ref{eq:il}), for
instance) up to 30 neighbors have to be taken into
account.\cite{Schnell2002} Depending on the specific band-structure
method used for the effective (auxiliary) one-particle problem, the
eigenfunctions (Bloch functions) are usually represented as linear
combinations of certain elementary functions, for instance plane waves,
Gaussians, or spherical harmonics.  Thus, also the evaluation of the
three-dimensional integral in (\ref{eq:onepartWan}) can usually be
traced back to known integrals over these elementary functions, and
usually at most a one-dimensional integral remains to be calculated
explicitly numerically.

The Coulomb matrix elements are given by:
\begin{equation} \label{eq:W1234}
  W_{12,34} = \int d^3\mr~ d^3\mr'~~ w^*_1(\mr) ~ w^*_2(\mr') ~~
  \frac{e^2}{|\mr-\mr'|} ~~ w_3(\mr') ~ w_4(\mr)
\end{equation}
Let us first take a brief look at general properties of the matrix
elements which are useful for minimizing computing time and memory storage.
From (\ref{eq:W1234}) and (\ref{eq:wtrans}), it follows that
\begin{equation}
  W_{12,34} = W_{(\mR_1-\mR_4n_1)(\mR_2-\mR_4n_2),(\mR_3-\mR_4n_3)(\mN n_4)}
\end{equation}
That is, we may always translate the lattice site indices in a way that
$\mR_4 \to \mN$.
Moreover, since $\mr$ and $\mr'$ in (\ref{eq:W1234}) can be interchanged,
we have $W_{12,34}=W_{21,43}$ and since $W$ is Hermitian
we have $W_{12,34}=W^*_{43,21}$.  
For the practical evaluation of these 6-fold integrals a fast Fourier
transformation (FFT) algorithm turned out to be very efficient and
independent of the specific representation of the Wannier functions.
Using 
\begin{equation} \label{eq:ft1r}
  \int d^3\mq~ \frac{e^{i\mq\mr}}{q^2} = \frac{2\pi^2}{|\mr|}
\end{equation}
one obtains:
\begin{equation} \label{eq:WM2}
  W_{12,34} = \frac{e^2}{2\pi^2} \int d^3\mq~
  \frac{1}{q^2} ~ f_1(\mq) ~ f_2(-\mq)
\end{equation}
where
\begin{equation} \label{eq:f1}
  f_1(\mq) = \int d^3\mr~ e^{i\mq\mr} ~
  w^*_{\mR_1n_1}(\mr) ~ w_{\mR_4 n_4}(\mr)
\end{equation}
\begin{equation} \label{eq:f2}
  f_2(\mq) = \int d^3\mr~ e^{i\mq\mr} ~
  w^*_{\mR_2n_2}(\mr) ~ w_{\mR_3n_3}(\mr)
\end{equation}
These functions are just the Fourier transforms of a product of
Wannier functions. They can be calculated very efficiently by evaluating the
Wannier functions on a cubic mesh in $\mr$-space with some finite spacing
$\Delta x$ and then applying a standard (numerical) FFT-algorithm.
The result of this Fourier transformation is $f_{1,2}(\mq)$
on a cubic mesh in $\mq$-space with some $\Delta q$.
As $F(\mq) = ~ f_1(\mq) ~ f_2(-\mq) $
is a smooth function at $\mq=\mN$, a divergence arises
in the integrand from $q^{-2}$.
In order to treat this divergence,
we split the integral by subtracting and adding $F(\mN)$:
\begin{equation} \label{eq:intFq}
  \int d^3\mq~ \frac{ F(\mq) }{q^2} =
  \int d^3\mq~ \frac{ F(\mq)-F(\mN) }{q^2} + F(\mN)
  \int d^3\mq~ \frac{1}{q^2}
\end{equation}
The first non-vanishing term of a polynomial expansion of the numerator
$F(\mq)-F(\mN)$ is of order $q^2$.
Hence, the polynomial expansion of the integrand starts with a constant
and the divergence is avoided.

All integrals are over a cube with length $2p=N\Delta q$.
The first integral in (\ref{eq:intFq}) is evaluated by
transforming the integral into a sum over little cubes with volume
$(\Delta q)^3$.
At $\mq=\mN$, the value of the integrand is calculated via the second
derivative of $F(\mq)$ at $\mq=\mN$ numerically.

The remaining integral in (\ref{eq:intFq}) is simply a constant given by:
\begin{eqnarray} \nonumber
  \int_{-p}^{+p}dq_x \int_{-p}^{+p}dq_y \int_{-p}^{+p}dq_z~ \frac{1}{q^2}
&=&
\\ \label{eq:mqw}
  p \int_{-1}^{+1}dq_x \int_{-1}^{+1}dq_y \int_{-1}^{+1}dq_z~ \frac{1}{q^2} 
&=& p \cdot C
\end{eqnarray}
with $C=15.34825$, which we have evaluated numerically.

As argued already for the one-particle matrix elements, the Coulomb
matrix elements need explicitly be evaluated only on-site, i.e. for
$\mR_1 = \mR_2 = \mR_3 = \mN$, and for at most only a few neighbor
shells, probably only nearest neighbors, because of the good
localization of the Wannier functions used. Here again the application
of the Marzari-Vanderbilt algorithm turns out to be important in order
to reduce the number of explicitly necessary evaluations of Coulomb
matrix elements, which are (computational) time consuming.

\section{\label{sec:manybody}Many-body treatment of
second quantized Hamiltonian with ab-initio parameters}

After the evaluation of the one-particle and two-particle matrix
elements we have the Hamiltonian in second quantization, i.e., in the
form (\ref{eq:secondquant}) in a Wannier representation with all
parameters determined from first principles. The only approximations
made so far are the restriction to a finite number $J$ of band indices
for a specific basis set and the approximations inherent to the
band-structure calculation method used (for example, muffin-tin
potentials, linearization approximations, discretization in $\mk$- and
real $\mr$-space, etc.). The Hamiltonian is in the form of a
multi-($J$-)band extended Hubbard model. A one-band description is
usually not sufficient even for the simplest solids. Furthermore, the
hopping matrix elements are in general not restricted to nearest
neighbors (but taken into account up to the 5th nearest neighbors) and
direct and exchange Coulomb matrix elements on-site, inter-band and
intra-band and, if necessary, also inter-site (but again only up to a
few neighbor shells) are taken into account. 

Now, in principle, one can apply any many-body method that has been
developed for interacting electron systems on a lattice. Many of the
standard methods rely on systematic perturbation theory with respect to
the Coulomb interaction, and the terms can be represented by Feynman
diagrams\cite{Mahan}. 
The lowest-order selfconsistent approximation within this diagrammatic
approach is the Hartree-Fock approximation (HFA), which
can also be derived from a variational principle. Within the HFA the
selfenergy is explicitly given by:
\begin{eqnarray} \label{eq:HF2q}
  \Sigma_{12,\sigma}^{\text{HF}} &=& \Sigma_{12}^{\text{Hart}}
                                   + \Sigma_{12,\sigma}^{\text{Fock}}
\\ \nonumber
  &=& \sum_{34\sigma'} \left[ W_{13,42} -
  \delta_{\sigma\sigma'}W_{31,42}\right] \bra c_{3\sigma'}^{\dagger}
  c_{4\sigma'} \ket
\end{eqnarray}
Here the expectation values (or the density matrix)
\begin{equation} \label{eq:densitym}
  A_{12}^\sigma = \bra c_{1\sigma}^{\dagger}c_{2\sigma} \ket
\end{equation}
have to be determined selfconsistently for the
Hartree-Fock Hamiltonian:
\begin{equation} \label{eq:HFHamilton}
  H_{\text{HF}} =
  \sum_{12\sigma} \left(t_{12} + \Sigma_{12,\sigma}^{\text{HF}}\right)
  c_{1\sigma}^\dagger c_{2\sigma}
\end{equation}
These expectation values are easily calculated by going back to the Bloch
states within which the effective one-particle 
Hamiltonian (\ref{eq:HFHamilton})
is diagonal with eigenenergies $E_m^\sigma(\mk)$. 
In this Bloch representation the expectation values for zero
temperature (i.e., within the ground state) are given by 
\begin{equation} \label{eq:densBloch}
  \bra c_{n\mk\sigma}^{\dagger} c_{n'\mk'\sigma'} \ket =
  \delta_{nn'}\delta_{\mk\mk'}\delta_{\sigma\sigma'} \theta(E_F -
  E_n^\sigma(\mk)) ~~.
\end{equation}
One finds
\begin{equation} \label{eq:densitym2}
  A_{12}^\sigma = \frac{1}{N} \sum_{\mk} e^{-i\mk(\mR_1 - \mR_2)} \sum_m
  U_{mn_1}^\mk U_{mn_2}^{\mk*} \theta(E_F - E_m^\sigma(\mk))
\end{equation}     
where (similar as in (\ref{eq:gauge})) the $U_{mn}^\mk$ denote the unitary 
transformation between the basis with band indices leading to
maximally localized Wannier functions and the basis within which the
effective one-particle Hamiltonian (\ref{eq:HFHamilton}) is diagonal.
For the total energy in the Hartree-Fock approximation one finds:
\begin{equation} \label{eq:HFenergy}
  E_{\text{HF}} = \bra H \ket = \sum_{12\sigma} \left( t_{12} +
  \frac{1}{2} \Sigma_{12}^{\text{Hart}} +
  \frac{1}{2} \Sigma_{12,\sigma}^{\text{Fock}}\right)
  A_{12}^\sigma
\end{equation}
This follows formally from the original many-body Hamiltonian 
(\ref{eq:secondquant}) using the decoupling
\begin{equation} \label{eq:HFdecoupling}
  \bra c_{1\sigma}^{\dagger}c_{2\sigma'}^{\dagger}c_{3\sigma'}c_{4\sigma} \ket
  \rightarrow \bra c_{1\sigma}^{\dagger}c_{4\sigma} \ket 
  \bra c_{2\sigma'}^{\dagger}c_{3\sigma'} \ket -
  \bra c_{1\sigma}^{\dagger}c_{3\sigma'} \ket 
  \bra c_{2\sigma'}^{\dagger}c_{4\sigma} \ket
\end{equation}
which is valid according to Wick's theorem, if the expectation values
are calculated with respect to an effective one-particle Hamiltonian,
which the Hartree-Fock-Hamiltonian (\ref{eq:HFHamilton}) is.

Of course the HFA is only the simplest and basic many-body approximation
which can be applied (mean-field). It is correct only up to linear order
in the Coulomb interaction. The main advantage of the diagrammatic
approach to the second-quantized many-body Hamiltonian
(\ref{eq:secondquant}) is that one can formulate systematic improvements
and corrections to the basic approximations. One possibility to go
beyond HFA is to take into account all skeleton diagrams up to second
order in the Coulomb interactions. Probably better and more reliable
many-body approximations are obtained by resummations of an infinite
series of diagrams of a certain class. For instance the resummation of
all bubble diagrams corresponds essentially to the random phase
approximation (RPA) and means that in the Fock diagram the bare
interaction has to be replaced by an effective screened interaction.

Of course the application of many-body theory is not restricted to the
standard perturbational methods in terms of Feynman diagrams. During the
last decade much progress in many-body theory has been made due to the
development and successful application of non-perturbative methods, such
as the DMFT\cite{GKKR96}. In its original version the DMFT requires a
local ($\mk$-independent) selfenergy, which is usually justified for
3-dimensional systems. But even if this assumption should not be
justified, the application of non-local (cluster) extensions of DMFT is
possible. In any case, for realistic materials a multi-band system has
to be studied; therefore a mapping on a multi-level single-impurity
problem and new types of DMFT-selfconsistency cycles are probably
required. Another possible non-perturbative many-body approach to our
Hamiltonian (\ref{eq:secondquant}) with ab-initio parameters is the
application of variational methods, such as the generalized Gutzwiller
ansatz, which has recently been applied to the 3d-ferromagnet
nickel\cite{GebhardWeber}.

\section{\label{sec:comparison}Comparison with other ab-initio methods}

In this section we compare our
ab-initio approach with existing and established
first-principles methods for electronic-structure calculations. 

DFT-based approaches encounter the problem that the density functional
of the exchange-correlation energy is not known, and hence additional
assumptions and ansatz are necessary. Within LDA the assumption is made
that the particle density to be determined is the same as that of an
effective one-particle system, and the ansatz for the (local) density
dependence of the exchange-correlation energy is chosen so that in the
case of the homogeneous electron gas the most accurately known results
are reproduced. So results of many-body theory  for the homogeneous
electron gas enter, but many-body theory is not directly applied for the
lattice system (an inhomogeneous electron gas). In LDA the (local)
Hartree potential for the inhomogeneous electron system is explicitly
taken into account when solving the Kohn-Sham equations, but the
exchange (Fock) contribution is only approximately considered (in a form
which is correct only for the homogeneous electron system). Therefore,
whereas self-interaction terms exactly cancel, if the Hartree and Fock
contributions are treated on the same level, they usually do not cancel
in LDA. Our approach, on the other hand, is free from the problem of
self-interaction, as can immediately be seen from
(\ref{eq:HFdecoupling}), because the Hartree- and exchange terms are
treated on the same level. Furthermore, many-body theory is directly
applied for the lattice electron system instead of only for the
homogeneous electron gas. From the very beginning we never assume a
constant electron density or a dependence on the local density. The
effects which the ``gradient correction'' and nonlocal density schemes
corrections to LDA aim at are automatically considered in our approach.   

The standard Hartree-Fock approximation in first quantization requires
the solution of the equations
\begin{eqnarray} \nonumber
\left( - \frac{\hbar^2}{2m} \nabla^2 + V(\mr)\right)\ph_i(\mr)
  + \sum_{j\sigma'} 
  \int d^3r' \frac{e^2|\ph_j(\mr')|^2}{|\mr - \mr'|}\ph_i(\mr)
\\ \label{eq:HFfirstquant} 
- \sum_j
  \int d^3r' \frac{e^2 \ph_j^*(\mr')\ph_i(\mr')}{|\mr - \mr'|} \ph_j(\mr) =
  \ep_i \ph_i(\mr) ~~.~~~~~~~~~
\end{eqnarray}
For equal spin and $i=j$ the second (Hartree) and the third (Fock or
exchange) contributions exactly cancel each other so that no
self-interaction occurs. The aim of the recent self-interaction
corrected (exact exchange) schemes\cite{Vogl97} as LDA improvements is,
therefore, to consider these exchange terms rigorously. But the
effective exchange (Fock) potential is a non-local potential, and the
one-particle wave-function $\ph_i$ to be determined enters not only at
the position $\mr$ but also at all other positions $\mr'$. For that
reason HFA treatments in first quantization and exact exchange schemes
are computationally more complicated than LDA methods, especially for a
periodic solid. Within our approach HFA-calculations are as simple as a Hartree
calculation, since only expectation values have to be evaluated. But in
our method self-interaction terms also exactly cancel. Therefore, our
approach should be as good as the exact-exchange approach but easier in
practical applications. 

When using an RPA (screened HFA) as the many-body approximation, our
approach should be similar to the GWA. But in GWA one usually
uses as the Green function (G-line) the LDA (effective one-particle)
result whereas we suggest using a selfconsistently calculated Green function
(within the many-body approximation used). Also, when using the
LDA Green function, only a part of the exchange is
taken into account unlike the exact exchange of our approximation. 

All the recent attempts to combine ab-initio and many-body methods (on
different levels concerning the many-body treatment), namely, the LDA+U,
the LDA++, and the LDA+DMFT-schemes, have in common that they start from
an LDA ab-initio calculation and obtain their one-particle bands and
density of states from this LDA treatment, and then add an interaction
(correlation) term with an on-site Hubbard-U term. Then, on one hand,
some of the correlations and interactions are already included in the
effective one-particle energies, namely those on the LDA (homogeneous
electron gas) level. On the other hand, since other correlations are
treated in many-body theory for the Hubbard U, there may be a double
counting of interaction (correlation) contributions, which is hard to
justify and the magnitude of which is difficult to estimate. Furthermore,
in these theories the Hubbard U usually is an additional free parameter,
so that these are not really ``ab-initio'' (first-principles)
approaches. In the our approach we do not start from the LDA
but from an effective band structure calculation on the Hartree level.
Therefore no correlation terms are implicitly included within the
one-particle band structures and the problem of double counting of
interaction terms does not occur. Furthermore, all Coulomb matrix
elements are calculated from ``first principles''.

\section{\label{sec:summary}Summary and Outlook}

In this paper we have suggested a new approach for combining ab-initio
and many-body methods for the calculation of the electronic properties of
solids. The starting point is a traditional band-structure calculation
for an effective (auxiliary) one-particle Hamiltonian, which can be the
Hartree-Hamiltonian. This yields, in particular, the eigenfunctions in
the form of Bloch functions. Keeping only a finite number of $J$ band
indices restricts and truncates the Hilbert space for further
calculations. We use the Marzari-Vanderbilt algorithm to construct
maximally localized Wannier functions (within the truncated one-particle
Hilbert space). Then all the one-particle (tight-binding) and
two-particle (Coulomb) matrix elements between these Wannier functions
can be calculated. The strong localization guarantees that only on-site
matrix elements and near-neighbor inter-site matrix elements have to be
calculated. We are left with a many-body Hamiltonian in second
quantization but with parameters determined from first principles for
any given material. This should be solved by a suitable many-body
approximation. The simplest approximation is the
(unscreened) HFA, but improved methods (e.g., RPA or non-perturbative
many-body approximations like DMFT) are also feasible. Our suggested
approach is free from the problems of double counting of
correlation effects and self-interaction and considers exchange
contributions exactly. It does not rely on assumptions based
on the homogeneous electron gas or a dependence on the 
local electron density density. An inhomogeneous (lattice) electron system is 
considered right from the beginning.

An application of this scheme to the 3d transition metals Fe, Co, Ni and Cu
using the unscreened HFA as the many-body method is presented in the 
following paper.\cite{Schnell2003-2} 

\begin{acknowledgments}
  This work has been supported by a grant from the Deutsche
  Forschungsgemeinschaft No. Cz/31-12-1.
  It was also partially supported by the Department of Energy
  under contract W-7405-ENG-36.  This research used resources of the
  National Energy Research Scientific Computing Center, which is
  supported by the Office of Science of the U.S. Department of Energy
  under Contract No. DE-AC03-76SF00098.
\end{acknowledgments}


\begin{thebibliography}{}
\bibitem[1]{HK64} P. Hohenberg, W. Kohn, Phys. Rev. {\bf 136},  B 864 (1964)
\bibitem[2]{DrGross90} R.M. Dreizler, E.K.U. Gross, {\sl Density Functional
  Theory}, Springer (Berlin, Heidelberg, New York 1990)
\bibitem[3]{Eschrig96} H. Eschrig, {\sl The Fundamentals of Density
  Functional Theory} (Teubner Stuttgart, Leipzig 1996)
\bibitem[4]{KS65} W. Kohn, L.J. Sham, Phys. Rev. {\bf 140}, A1133 (1965)
\bibitem[5]{Rose84} J.H. Rose, J.R. Smith, F. Guinea and J. Ferrante,
  Phys. Rev. B {\bf 29}, 2963 (1984).
\bibitem[6]{DrGrossKap7} Chapter 7 of Ref. \onlinecite{DrGross90}
\bibitem[7]{Gunnarsson80} O. Gunnarsson, R.O. Jones,
  Physica Script. {\bf 21}, 394 (1980)
\bibitem[8]{Vogl97} M. Stadele, J.A. Majewski, P.Vogl, A. Gorling, Phys. Rev.
  Letters {\bf 79}, 2089 (1997)
\bibitem[9]{Hedin} L. Hedin, B.I. Lundquist, Solid State Physics {\bf 23},
  p.1 (Eds.: F. Seitz, D. Turnbull, H. Ehrenreich, Academic Press 1969)
\bibitem[10]{Aulbur} For a recent review see: W. Aulbur, L. J{\"o}nsson,
  J.W. Wilkins in: Solid State Physics {\bf 54}, p.2 (Eds.: H. Ehrenreich,
  F. Spaepa, Academic Press 2000)
\bibitem[11]{AZA91} V.I. Anisimov, J. Zaanen, O.K.Andersen,
  Phys. Rev. B {\bf 44}, 943 (1991)
\bibitem[12]{SASS91} M.M. Steiner, R.C. Albers, D.J. Scalapino, L.J. Sham,
  Phys. Rev. B {\bf 43}, 1637 (1991)
\bibitem[13]{SAS9294} M.M. Steiner, R.C. Albers, L.J. Sham,
  Phys. Rev. B {\bf 45}, 13272 (1992); Phys. Rev. Lett. {\bf 72}, 2923 (1994)
\bibitem[14]{APKAK97} V.I. Anisimov, A.I. Poteryaev, M.A. Korotin, A.O.
  Anokhin, G. Kotliar, J. Phys. Cond. Matter {\bf 9}, 7359 (1997)
\bibitem[15]{LK98} A.I. Lichtenstein, M.I. Katsnelson, Phys. Rev. B {\bf 57},
  6884 (1998)
\bibitem[16]{DJK99} V. Drchal, V. Janis, J. Kudrnovsky, Phys. Rev. B {\bf
  60}, 15664 (1999)
\bibitem[17]{KL99} M.I. Katsnelson, A.I. Lichtenstein,
  J. Phys. Cond. Matter {\bf 11}, 1037 (1999)
\bibitem[18]{LL00} A. Liebsch, A. Lichtenstein, Phys. Rev. Lett. {\bf 84},
  1591 (2000)
\bibitem[19]{WPN00} T. Wegner, M. Potthoff, W. Nolting,
  Phys. Rev. B {\bf 61}, 1386 (2000)
\bibitem[20]{NHBPAV00} I.A. Nekrasov, K. Held, N. Bl{\"u}mer, A.I.
  Poteryaev, V.I. Anisimov, D. Vollhardt, Eur. Phys. J. B {\bf 18}, 55 (2000)
\bibitem[21]{GKKR96} For a review see: A. George, G. Kotliar, W. Krauth,
  M.J. Rozenberg, Rev. Mod. Phys. {\bf 68}, 13 (1996)
\bibitem[22]{HeldVollhardtetal} K. Held, I.A. Nekrasov, G. Keller, V. Eyert,
  N. Bl{\"u}mer, A.K. McMahan, R.T. Scalettar, T. Pruschke, V.I. Anisimov,
  D. Vollhardt, in: {\sl Quantum Simulations of Complex Many-Body systems: From
  Theory to Algorithms}, p.175, (eds.: J. Grotendorst, D. Marx, A. Muramatsu,
  NIC Series Vol. 10, Forschungszentrum J{\"u}lich 2002)
\bibitem[23]{MV97} N. Marzari and D. Vanderbilt, Phys. Rev. B {\bf 56},
  12847 (1997).
\bibitem[24]{OKAndersen} O.K. Andersen, Phys. Rev. B {\bf 12}, 3060 (1975)
\bibitem[25]{Skriver} H.L. Skriver,  {\em The LMTO Method}
  (Springer-Verlag, Heidelberg 1984)
\bibitem[26]{Singh94} D. Singh, {\sl Plane waves, pseudopotentials and the
  LAPW method} (Kluwer Academic, Amsterdam 1994)
\bibitem[27]{WIEN2K} P. Blaha, K. Schwarz, G. Madsen, D. Kvasnicka, J.
  Luitz, {\sl WIEN2k, An Augmented Plane Wave + Local Orbitals
  Program for Calculating Crystal Properties} (Karlheinz Schwarz,
  Techn. Universit{\"a}t Wien, Austria), 2001. ISBN 3-9501031-1-2
\bibitem[28]{Eschrig99} K. Koepernik and H. Eschrig, Phys. Rev. B {\bf 59},
  1743 (1999);
  I. Opahle, K. Koepernik, and H. Eschrig, Phys. Rev. B {\bf 60}, 14035
  (1999).
\bibitem[29]{MV2002} I. Souza, N. Marzari, and D. Vanderbilt, Phys. Rev. B
  {\bf 65}, 035109 (2002)
\bibitem[30]{Scalettar2002} W. Ku, H. Rosner, W. E. Pickett, and R. T.
  Scalettar, Phys. Rev. Lett. {\bf 89}, 167204 (2002)
\bibitem[31]{Schnell2002} I. Schnell, G. Czycholl, R.C. Albers,
  Phys. Rev. B {\bf 65}, 075103 (2002)
\bibitem[32]{Mahan} G.D. Mahan, {\sl Many-Particle Physics} (Plenum Press New
  York 1990)
\bibitem[33]{GebhardWeber} W. Weber, J. B{\"u}nemann, F. Gebhard, in{\sl
  Band-Ferromagnetsim} (Lecture Notes in Physics, Vol. 580, p.9 (eds.: 
  K.Baberschke, M. Donath, W.Nolting, Springer Berlin 2001); J. B{\"u}nemann,
  W. Weber, F. Gebhard, Phys. Rev. B {\bf 57}, 6896 (1998)
\bibitem[34]{Schnell2003-2} I. Schnell, G. Czycholl, R.C. Albers, Phys. Rev. B
  (2003), following paper 
\end{thebibliography}
\end{document}